\documentclass[letterpaper,10pt]{article} 

\usepackage{opticameet3} 

\newcommand\authormark[1]{\textsuperscript{#1}}

\usepackage[colorlinks=true,bookmarks=false,citecolor=blue,urlcolor=blue]{hyperref} 
\usepackage{amsmath,amssymb}
\usepackage{cleveref}
\crefname{equation}{Eq.}{Eq.} 
\crefrangelabelformat{equation}{(#3#1#4)--(#5#2#6)}
\crefname{figure}{Fig.}{Fig.}
\crefname{table}{Table}{Table}

\begin{document}

\title{Experimental Demonstration of End-to-End Optimization for Directly Modulated Laser-based IM/DD Systems}


\author{Sergio Hernandez\authormark{1}, Christophe Peucheret\authormark{2}, Francesco Da Ros \authormark{1} and Darko Zibar \authormark{1}}

\address{\authormark{1} Dep. of Electrical and Photonics Eng., Technical University of Denmark, 2800 Kongens Lyngby, Denmark\\
\authormark{2} Univ. Rennes, CNRS, FOTON - UMR6082, 22305 Lannion, France}
\email{shefe@dtu.dk} 

\begin{abstract}
We experimentally demonstrate the joint optimization of transmitter and receiver parameters in directly modulated laser systems, showing superior performance compared to nonlinear receiver-only equalization while using fewer memory taps, less bandwidth, and lower radiofrequency power.
\end{abstract}
\vspace{-0.05cm}
\section{Introduction}
Directly modulated lasers (DMLs) offer a promising solution for intensity modulation / direct detection (IM/DD) short-reach optical communication systems thanks to to their low power consumption, compact size, and affordability \cite{Matsuo:18}. Recent works have demonstrated DML modulation bandwidths beyond 100 GHz, allowing data rates of more than 250~Gbps \cite{Yamaoka2021}. However, the response of DMLs is governed by nonlinear differential rate equations, that deviate from the traditional additive white Gaussian noise (AWGN) framework. The complex response of DMLs makes the determination of the optimal pulse and constellation shaping challenging, but it also prevents the propagation of gradients between transmitter (TX) and receiver (RX), hindering the application of gradient-based optimization techniques. 
Prior work on impairment compensation \cite{Matsuo:18} in DML systems has focused on RX-side equalization (EQ) or the alternating optimization of digital pre-distortion (DPD) and EQ filters \cite{Kottke:17}, where the propagation of gradients between TX and RX is not necessary. End-to-end (E2E) learning is an optimization strategy that involves adjusting system parameters at TX (constellation and pulse shaping) and RX (equalizer, detection) jointly, potentially leading to optimal digital signal processing (DSP) configurations \cite{Minelli23}. Although gradient-free and online optimization methods can avoid the need for gradient propagation, differentiable surrogate models can enable it while being less sensitive to noise and potentially reducing the required number of system propagations.

In this paper, we extend our previous numerical work \cite{Hernandez:24} to experimentally demonstrate the E2E optimization of a DML-based system using an autoencoder (AE). The proposed configuration includes the optimization of the TX and RX DSP, the bias current $I_\mathrm{bias}$, and the power $P_\mathrm{RF}$ of the radiofrequency (RF) modulating signal. It is applied to optimizing a 4-pulse amplitude modulation (4PAM) DML-based back-to-back system operating at 20 and 30 GBaud. The optimization utilizes a pre-trained surrogate DML model based on experimental data and a long short-term memory (LSTM) network, building upon our previous work on DML modeling \cite{US:ECOC23}. The performance of the AE is compared to two RX-side equalization configurations (RX-EQ): one based on a linear finite impulse response (FIR) feedforward equalizer (FFE) and a second one based on Volterra nonlinear equalization (VNLE). The results show that the AE system can deliver better error performance than its counterparts while requiring fewer filter taps and lower input modulation amplitude, thanks to an improved bandwidth (BW) utilization and nonlinearity mitigation.

\section{Experimental setup}
\begin{figure}[t]
    \centering
    \includegraphics[width=\linewidth]{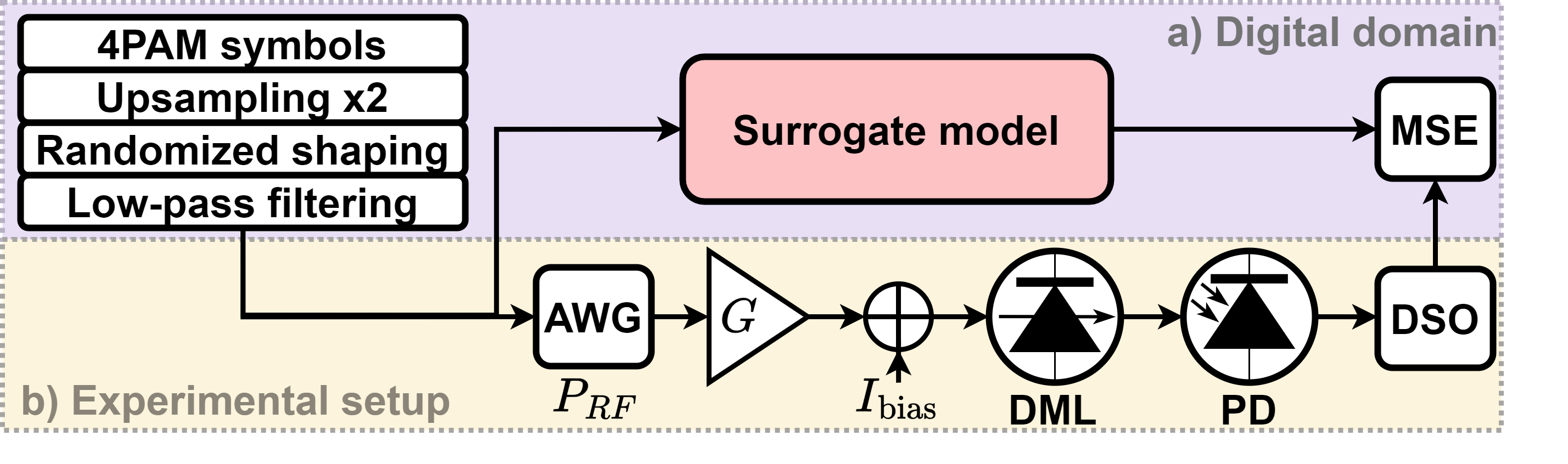}
    \caption{Block diagram of the proposed DML surrogate modeling setup. The model is trained on discrete input/output sequences (a) obtained by propagation through the experimental setup (b).}
    \label{fig:setup}
\end{figure}

The proposed optimization of the AE configuration consists of two stages: the surrogate modeling of the experimental setup and the optimization of the system based on the obtained surrogate. The experimental setup consists of an arbitrary waveform generator (AWG) (BW:~25~GHz), an RF amplifier (BW:~38~GHz) with a gain $G = 13$~dB, a DC power supply generating $I_{\mathrm{bias}}$, an NTT NLK1551SSC DML (10~Gbps class, external efficiency:~0.15~W/A) controlled in temperature to 25°C, a photodetector (PD) (BW:~50~GHz) and a digital sampling oscilloscope (DSO) (BW:~33~GHz). The surrogate training dataset consists of input and output waveforms derived from propagating digital sequences through the experimental setup, depicted in \cref{fig:setup}. The digital sequences are produced by generating equiprobable 4PAM symbols and applying randomized pulse shaping to them \cite{US:ECOC23}. The bias current $I_{\mathrm{bias}} \in [50, 100]$ mA and $P_\mathrm{RF} \in [-4, 2]$ dBm are also randomized. This allows the model to capture the DML dynamics under a wide range of input waveforms and power ranges. The input digital sequences are upsampled to the AWG sampling rate (65~GSa/s) and low-pass filtered with a 9-tap super-Gaussian FIR filter (2nd order) before digital to analog conversion to match the AWG BW. After capturing the output waveform with the DSO 25 times, the obtained sequence is downsampled, temporally synchonized with the input sequence, averaged (turning the 25 copies into a single denoised one) and normalized to the range $[0,1]$. The LSTM surrogate is then iteratively trained using mean squared error (MSE) between the model output and the output sequences from the DSO using an Adam optimizer to update the model weights. The obtained testing MSE was $1.84\times 10^{-4}$ and $1.75\times 10^{-4}$ for $R_{s} =$ 20 and 30 GBaud, respectively. Once the surrogate model is trained, its weights are fixed and it is used as part of the E2E system optimization to train the AE configuration.

\vspace{0cm}

\begin{figure}[hb!]
\centering
\includegraphics[width=\linewidth]{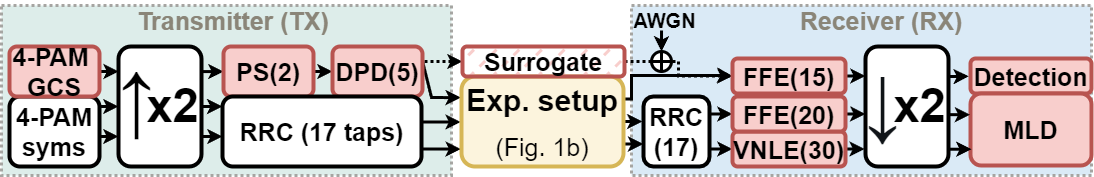}
\caption{Block diagram of the 3 presented configurations. The AE configuration (top) is trained using the surrogate model (dotted arrow) and tested on the experimental setup. The optimized blocks are highlighted in red, and the numbers in parenthesis represent the number of memory taps of the block.}
    \label{fig:e2esetup}
\end{figure}

The DSP structures of the three investigated configurations (AE, RX-only linear EQ, RX-only nonlinear EQ) are shown in \cref{fig:e2esetup}. All three configurations are based on equiprobable 4PAM symbols, upsampled to 2 samples per symbol (SpS). The AE configuration (top of the figure) consists of geometric constellation shaping (GCS), 2-tap discrete-time FIR pulse shaping, and a 5-tap DPD FIR filter to generate the input sequences on the TX side. The AE sequences are low-pass filtered (using the aforementioned FIR super-Gaussian filter) to match the AWG BW before conversion to the analog domain. The AE configuration must also determine the optimal $I_{\mathrm{bias}}$ and $P_\mathrm{RF}$ within their respective allowed ranges. During optimization, the emulation of the system response is carried out by the surrogate model, as its architecture allows for automatic differentiation (i.e. gradient propagation). The testing is conducted on the experimental setup, allowing direct comparison with the other DSP configurations. Given that the deterministic surrogate does not introduce additive noise to the signal, AWGN noise is added after propagation through the surrogate. The noise variance is adjusted based on a signal-to-noise ratio (SNR) estimation from experimental data, using the 25 waveform copies from the surrogate dataset to distinguish between the deterministic and stochastic components of the signal for $P_{RF} = 2$~dBm. On the RX side, the AE performs 2-SpS-wise FFE, downsampling and symbol detection. The detection is performed using the normalized exponential function (Softmax) on the estimated symbol probabilities provided by the AE. For the two RX-EQ configurations, 17-tap root raised cosine (RRC) pulse shaping with roll-off factor $\alpha = 0.1$ is performed at the TX, complemented by a matched filter on the RX side. The difference between the two is in the equalization approach: the linear EQ uses a 20-tap FIR FFE filtering, while the nonlinear EQ uses a VNLE with 20 first-order taps and 10 second-order taps. Symbol detection is performed using maximum likelihood detection for both the FFE and VNLE configurations. 


\vspace{-0.15cm}

\section{Results}
\begin{figure}[t]
    \centering
    \includegraphics[width=\linewidth]{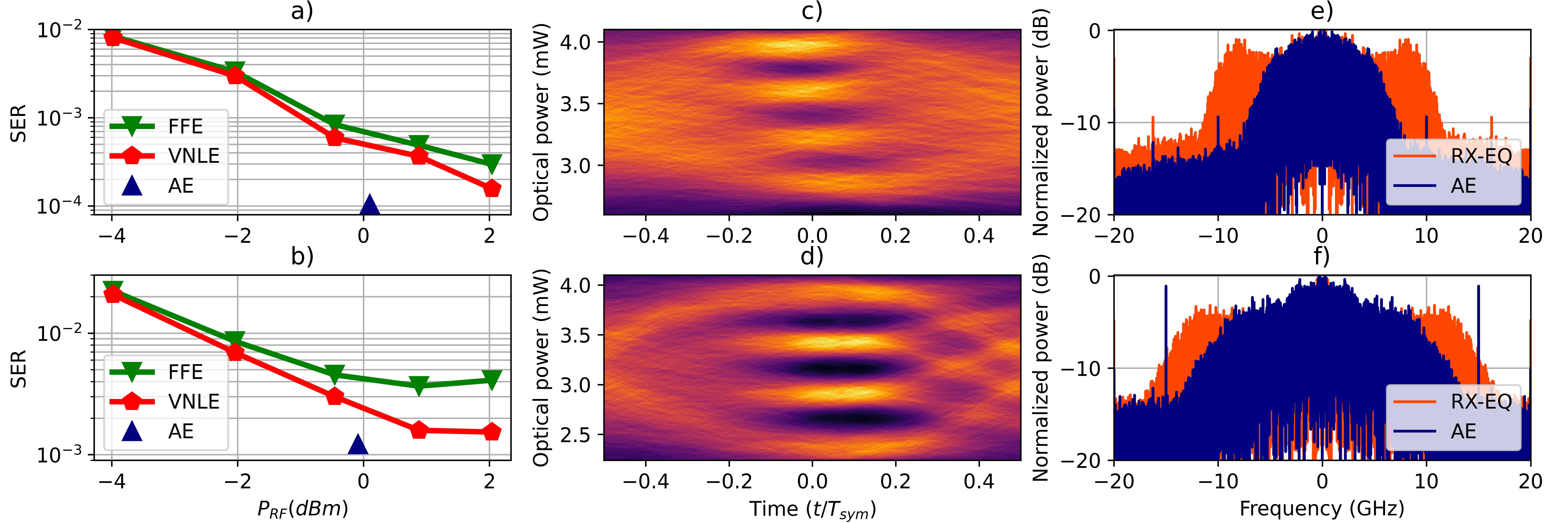}
    \caption{SER performance at 20~GBaud (a) and 30~GBaud (b); eye diagrams at for the 20~GBaud RX-EQ (c) and AE (d) signals before RX DSP; RX-EQ and AE spectra from the DSP at 20~GBaud (e) and 30~GBaud (f).}
    \label{fig:results}
\end{figure}

Given that the AE system can optimize $I_{\mathrm{bias}}$ and $P_\mathrm{RF}$, the RX-EQ configurations were tested on the $I_{\mathrm{bias}}$ learned by the AE model (52.6 mA at 20 GBaud and 56 mA at 30 GBaud) while sweeping $P_\mathrm{RF}$ between -4 and 2~dBm, i.e. the power range allowed for the AE system during training. The symbol error rate (SER) performance of each system is shown for 2 symbol rates $R_{s} = \{20, \; 30\}$ GBaud in \cref{fig:results}(a) and (b), respectively. The AE achieves the best SER performance at both $R_{s}$, despite using a $P_\mathrm{RF}$ of around 0 dBm, lower than the maximum value of 2~dBm. The performance difference between the FFE and the VNLE is significantly higher at 30~GBaud, hinting at a higher nonlinear system response at higher $R_s$. The eye diagrams of the RX-EQ and AE sequences from the DSO at 20~GBaud (\cref{fig:results}(c) and (d), respectively) show the improved performance more visually. The higher optical modulation amplitude of the AE translates into a higher SNR to the AE signal, minimizing error probability and facilitating the equalization task on the RX side, as shown by the vertical opening of the AE diagram. Another interesting insight is the visible eye skew in the RX-EQ diagram. Eye skew can make the system more prone to timing errors, degrading error performance. The AE signal does not show such a skew, probably due to the effect of GCS and DPD on the TX side. Given that the AE configuration is able to tune the pulse shaping and DPD filters, the spectra of the output signal is also of interest. \cref{fig:results}(e) and (f) show the spectra of the RX-EQ and AE signals at 20 and 30~GBaud. It becomes apparent how the BW of the AE signal is significantly narrower than that of the RRC at both $R_{s}$, although the difference is more prominent in the 20 GBaud case. 
The AE is able to substantially compress the BW signal (49\% BW compression at 20~GBaud and 17\% at 30 GBaud, considering the -10 dB BW). The time and frequency domain representation of the output signals suggests that the AE is employing advanced signaling to reduce BW requirements while introducing a controlled amount of intersymbol interference (ISI). It should be noted that the AE system was not preconfigured for this purpose and the surrogate was trained only with 2-SpS waveforms. This highlights the AE's ability to adapt to the system's physical constraints, enabling it to achieve parameter configurations that would be challenging to obtain through conventional methods.

\section{Conclusion}
We have experimentally demonstrated the use of end-to-end learning toward the optimization of directly modulated laser-based systems based on a differentiable, data-driven surrogate model. The proposed autoencoder approach achieves better error performance than receiver-side equalization approaches while using 2~dB lower modulation radiofrequency power and reducing the bandwidth utilization of the system by more than 15\%.

\section{Acknowledgements} The Villum VI-POPCOM (no. VIL5448) and Villum YIP OPTIC-AI (no. VIL29344) projects are acknowledged.

\bibliographystyle{IEEEtran} 
\bibliography{refs}

\end{document}